\documentclass[aps,amsmath,amssymb,prb,twocolumn,showpacs,superscriptaddress]{revtex4-1}
\usepackage{bm}
\usepackage{epsfig}
\usepackage[usenames]{color}
\usepackage{graphicx}
\usepackage[hypertex]{hyperref}
\DeclareMathOperator{\Imm}{Im}

\DeclareMathOperator{\D}{\mathcal D}

\renewcommand{\paragraph}[1]{\textit{#1.---} }

\newcommand{\Gin}{\overrightarrow{\Gamma}^{(\mathrm{LR})}}
\newcommand{\Gout}{\overleftarrow{\Gamma}^{(\mathrm{RL})}}

\begin{document}
\title{Transport characteristics of nanojunctions far-from-equilibrium}

\author{A.~Glatz}
\affiliation{Materials Science Division, Argonne National Laboratory, Argonne, Illinois 60439, USA}

\author{N.~M.~Chtchelkatchev}
\affiliation{Institute for High Pressure Physics, Russian Academy of Science, Troitsk 142190, Russia}
\affiliation{Department of Theoretical Physics, Moscow Institute of Physics and Technology, 141700 Moscow, Russia}

\author{I.~S.~Beloborodov}
\affiliation{Department of Physics and Astronomy, California State University Northridge, Northridge, CA 91330, USA}

\date{\today}

\begin{abstract}
We study the tunneling transport through a nanojunction in the far-from-equilibrium regime at relatively low temperatures. We show that the current-voltage characteristics is significantly modified as compared to the usual quasi-equilibrium result by lifting the suppression due to the Coulomb blockade.
These effects are important in realistic nanojunctions. We study the high-impedance case in detail to explain the underlying physics and construct a more realistic theoretical model for the case of a metallic junction taking into account dynamic Coulomb interaction. This dynamic screening further reduces the effect of the Coulomb blockade.
\end{abstract}

\maketitle

\section{Introduction}

Great efforts in contemporary materials science research focus on transport properties of advanced nano-materials. 
In particular in arrays of nano particles, the interest is motivated by the fact that these can be treated as artificial solids with programmable electronic properties~\cite{beloborodov+rmp07}. The ease of adjusting electronic properties of granular materials is one of their most attractive assets for fundamental studies of disordered solids and for targeted applications in nanotechnology.  The parameters of granular materials are in many ways  determined by the properties of their building blocks: grains and tunnel junctions. The equilibrium properties of single grains and single junctions are well understood~\cite{Grabert_ch1}. However, much less is known about the far-from-equilibrium properties of those systems, by which we mean that the system properties cannot be described by just a perturbed equilibrium (or quasi-equilibrium) considerations. The understanding of far-from-equilibrium effects in tunnel junctions, the building blocks of most advanced nano-materials, is especially important for practical applications. Indeed, recent experimental research has focused on instabilities in the current-voltage characteristics showing clear deviations from quasi-equilibrium results~\cite{cohen+prb11,ovadia+prl09}. This defines an urgent quest for a quantitative description of far-from-equilibrium properties of a single tunnel junction. 

\begin{figure}[b]
  \includegraphics[width=0.9\columnwidth]{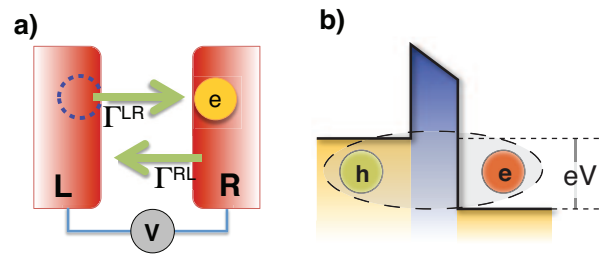}\\
  \caption{(color online) a) Sketch of the tunnel junction with two leads and b) illustration of the electron-hole pair generation across the tunnel junction. These virtual bosonic excitations together with the probability to interact with the environmental modes inside the junction determines the total tunneling current. }\label{fig.env}
\end{figure}

In this paper we investigate the far-from-equilibrium current-voltage characteristics of a tunnel junction (see Fig.~\ref{fig.env}a). Electron transport in tunnel junctions is ensured by the energy exchange between the tunneling electrons and energy reservoirs: since the electronic energy levels at the leads are unequal, tunneling is only possible if a subsystem of excitations capable of accommodating this energy difference exists. At not very high temperatures, where the phonon density is small, the role of the energy reservoir is played by an electromagnetic environment comprised of electron-hole pairs self-generated by the tunneling electrons. Here we concentrate on this low-temperature situation where phonons (bath) are irrelevant for the tunneling transport. In our approach the interaction time between electrons and environment needs to be much smaller than the one between environment and bath, in order to have a fully developed environment.

\section{Physical description of non-equilibrium effects in a nanojunction}

\begin{figure}[t]
\includegraphics[width=0.95\columnwidth]{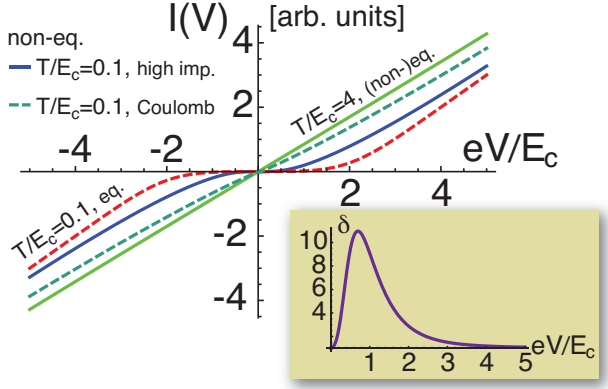}\\
  \caption{(color online) Current-voltage characteristics of nanojunctions at low temperatures ($T/E_c=0.1$, with $E_c$ being the Coulomb energy of the nanojunction) for the equilibrium case ``eq.'' (red, dashed curve, see also [\onlinecite{devoret+prl90}]) and the non-equilibrium case ``non-eq.'', which shows a clear enhancement of the current and reduced suppression due to Coulomb blockade at low voltages. This is in particular clear in the plot of the ratio $\delta=I_{\rm non-eq., high imp.}(V)/I_{\rm eq.}(V)$ in the inset. The linear (green, solid) curve shows the high temperature ohmic regime, which is independent of non-eq. or eq. considerations. The two non-eq. curves correspond to the high impedance case (blue, solid) and the dynamic Coulomb interaction case (dashed, turquois); see text for detailed explanations. }\label{fig.IVcomp}
\end{figure}

We start by expressing the tunneling current through a single junction as the difference of the electrons going from the left [L] to the right [R] electrode and the ones traversing the junction from right to left~\cite{Grabert_ch1,Grabert_ch2} (see also Fig.~\ref{fig.env}a)
\begin{equation}\label{eq:I}
    I = e\left[\Gin - \Gout\right].
\end{equation}
Here the tunneling current obeys the symmetry $I(-V) = - I(V)$ and
$\Gin$ $\left[\Gout\right]$ is the tunneling rate from the left (right) to the right (left)
 \begin{eqnarray}
        \Gin &=& \frac{1}{R_{\scriptscriptstyle {\mathrm T}}}
        \int\limits_{-\infty}^{+\infty} \int\limits_{-\infty}^{+\infty} d\varepsilon d\varepsilon' f_\varepsilon^{(L)}
        (1-f_{\varepsilon'}^{(R)})P(\varepsilon-\varepsilon' + eV)\label{eq:Gin} \nonumber \\
        &=&\frac{1}{R_{\scriptscriptstyle {\mathrm T}}}\int\limits_{-\infty}^{+\infty}d\varepsilon\,(\varepsilon-eV)N_{\varepsilon}(-eV)P(\varepsilon) ,
\end{eqnarray}
where $f_\varepsilon^{(L,R)}$ are the electronic distribution functions within the leads, $R_T$ is the tunnel resistance, and $V$ is the voltage difference across the junction. The terms $f_\epsilon^{(L)}$ and $(1-f_{\epsilon'}^{(R)})$ correspond to the occupied electron state with energy $\epsilon$ in the left lead and the hole state with
energy $\epsilon'$ in the right lead, respectively. 
The function $P(\varepsilon)$ determines the probability that the tunneling electron loses [gains] the energy $\varepsilon$ to [from] environment modes in the junction. In general, both $f_\varepsilon^{(L,R)}$ and $P(\varepsilon)$ are out-of-equilibrium functions.
In the second line of Eq.~(\ref{eq:Gin}) we introduced a bosonic distribution function, $N_{\varepsilon}(eV)$, which describes an electron-hole excitations \textit{across} the junction, Fig.~\ref{fig.env}b). Explicitly it is given by $N_{\varepsilon}(eV)\equiv \left(\varepsilon+eV\right)^{-1}\int_{-\infty}^{+\infty}d\omega\,f_{\omega+(\varepsilon+eV)/2}^{(L)}\left(1-f_{\omega-(\varepsilon+eV)/2}^{(R)}\right)$. If the distribution functions at the electrodes are Fermi functions with equal temperatures $T$, then $N_\varepsilon (eV) = N_B(\varepsilon+eV,T)$, with $N_B(\varepsilon, T)$ being the equilibrium Bose distribution function. 

The corresponding backward rate in Eq.~(\ref{eq:I}) is $\Gout=  \frac{1}{R_{\scriptscriptstyle {\mathrm T}}} \int d\varepsilon d\epsilon' \left(1-f_{\epsilon}^{(L)}\right)f_{\varepsilon'}^{(R)}P(\varepsilon' -\epsilon - eV)=\frac{1}{R_{\scriptscriptstyle {\mathrm T}}}\int\limits_{-\infty}^{+\infty}d\varepsilon\,(eV-\varepsilon)N_{-\varepsilon}(eV)P(-\varepsilon)$. Here we note, that the bosonic distribution function $N_\varepsilon(eV)$ depends in general on both lead temperatures; and the bosonic form of the backward rate $\Gout$ can be written in this form only if $f_\varepsilon^{(L)}$ and $f_\varepsilon^{(R)}$ have the same functional dependence on energy and temperature.

Using the bosonic description of the environment, the current-voltage characteristics Eq.~(\ref{eq:I}) can be written explicitly as 
\begin{equation}\label{eq:Ibos}
I = e \int_{-\infty}^{+\infty} d\varepsilon\,(\varepsilon-eV)\left\{N_\epsilon(-eV) P(\varepsilon) +N_{-\varepsilon}(eV) P(-\varepsilon)\right\}.
\end{equation}

This expression is general and we only assumed that the electron distribution functions in the contact leads have the same functional dependence on temperature and energy. In the out-of-equilibrium situation the tunneling electrons also interact with the environment {\it inside} the junction. This environment is self-generated by the tunneling electrons and thermal fluctuations, and its influence on the tunneling transport is implicitly taken into account through the probability function $P(\varepsilon)$ in Eq.~(\ref{eq:Ibos}). In the case of heat transport, the situation is different, since heat can dissipate inside the junction, see Ref.~[\onlinecite{glatz+prb11}] and the environment needs to be taken into account explicitly, having its own distribution function $n_{\rm env}(\varepsilon,eV)=\left[(\varepsilon-eV)N_\epsilon(-eV)+(\varepsilon+eV)N_\epsilon(eV)\right]/(2\varepsilon)$, Refs.~[\onlinecite{glatz+prb11},\onlinecite{chtch+prl09}].

As mentioned before, if the leads are in equilibrium (which is mostly the case due to their bulk nature) and in addition the temperatures at the leads are the same, $N_\epsilon(eV)$ in Eq.~(\ref{eq:Ibos}), becomes a Bose distribution function. In the following we assume that this is the case, but emphasize that the junction environment is always far-from-equilibrium if a finite voltage is applied.

Besides the distribution function $N_\epsilon(eV)$, the main distinguishing feature of our out-of-equilibrium consideration is the presence of the probability function $P(\varepsilon)$, which takes into account the interaction with the environment, determining the probability for the tunneling electrons to exchange the energy $\varepsilon$, which is the excess energy of the electrons compared to the potential difference of the leads. It is clear that this probability should decay for large energies and have a maximum when the energy matches the energy at which the environment resonates.
A second observation is that at large voltages this probability should get smeared out and the current is mostly determined by the distribution function $N_\varepsilon(eV)$ rather than $P(\varepsilon)$ which therefore determines only the resistance of the junction in the Ohmic regime.
Furthermore, high temperatures also broaden the probability. Both effects can be conveniently described by the introduction of an effective electron temperature $T_e= (eV/2) \coth(eV/(2T))$ [derived from $T_e=\lim_{\varepsilon\to 0} n_{\rm env}(\varepsilon,eV)$], which is equal to $T$ for small voltages and proportional to $eV/2$ for $eV\gg T$.~\cite{glatz+prb11,chtch+prl09}
Therefore, the probability function $P(\varepsilon)$ can be approximated by a Gaussian function where the electron temperature  $T_e$ determines its width in the high resistive case, which we will discuss in detail in the next section.

\section{Nanojunction with high impedance environment}
\label{highimpedance}

\begin{figure}[t]
\includegraphics[width=0.95\columnwidth]{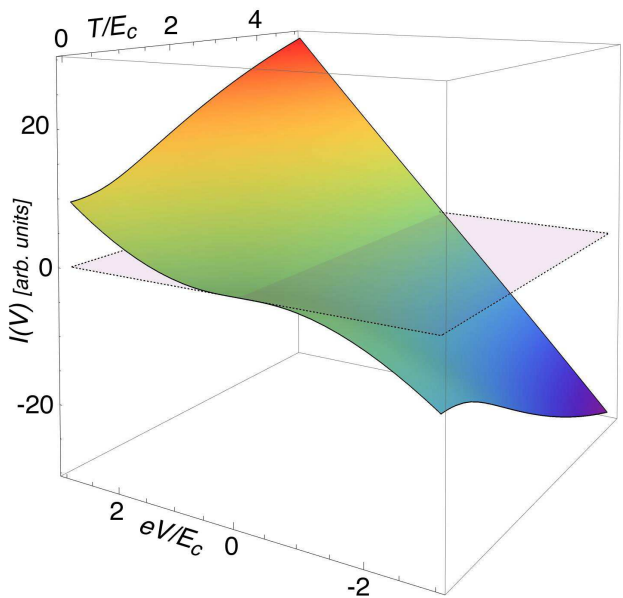}\\
  \caption{(color online) Far-from-equilibrium current-voltage characteristics of nanojunction depending on voltage and junction temperature. 
  Here $E_c$ is the Coulomb energy of the nanojunction. One clearly sees the crossover from the suppressed current at small voltages and temperatures to the Ohmic regime at high temperatures.}\label{fig.3DIV}
\end{figure}

We now turn to the experimentally important case of an environment with a
high impedance as compared to the quantum resistance, $R_{\mathrm{Q}}$.
In this limit, the tunneling electrons easily excite the environment modes.
The probability function $P(\epsilon)$ for electron-hole pairs with energy $\epsilon$ to appear in the junction 
in Eq.~(\ref{eq:Ibos}) can be written as 
\begin{equation}\label{eq:P1}
P(\varepsilon) = (1/\sqrt{2\pi \Delta^2}) \exp \left[- (\varepsilon - 2E_c)^2 /2\Delta^2\right].
\end{equation}
Here $\Delta = 2(E_c T_{\rm e})^{1/2}$ is the characteristic width of the distribution function with $E_c$ being the Coulomb energy of the nano tunnel junction. We note that this form of the probability function $P(\epsilon)$ in Eq.~(\ref{eq:P1}) depends on the electron temperature $T_e$ and not on the lead temperature $T$, as in the quasi-equilibrium case~\cite{devoret+prl90}.

Substituting this function $P(\epsilon)$ into  Eq.~(\ref{eq:Ibos}), we obtain our first
main result for the current-voltage characteristics of a tunnel junction. In particular, Fig.~\ref{fig.IVcomp} represents the $I-V$ 
characteristics at low temperatures ($T/E_c=0.1$) for the equilibrium case ``eq.'' (red, dashed curve, see also Ref.~\onlinecite{devoret+prl90}) and the non-equilibrium case ``non-eq.''. It shows a clear enhancement of the current and reduced suppression due to Coulomb blockade at low voltages. This is in particular clear in the plot of the ratio $\delta=I_{\rm non-eq.}(V)/I_{\rm eq.}(V)$ in the inset. The  (green, solid) linear curve shows the high temperature ohmic regime, which is independent of non-eq. or eq. considerations.  The full temperature and voltage dependence of the current-voltage characteristics of tunnel junction is shown in Fig.~\ref{fig.3DIV}. This figure clearly shows the crossover from the suppressed current at small voltages ($eV \ll E_c$) and temperatures ($T \ll E_c$) to the ohmic regime at high temperatures.

\section{Nanojunction with dynamic Coulomb interaction}

\begin{figure}[t]
\includegraphics[width=0.95\columnwidth]{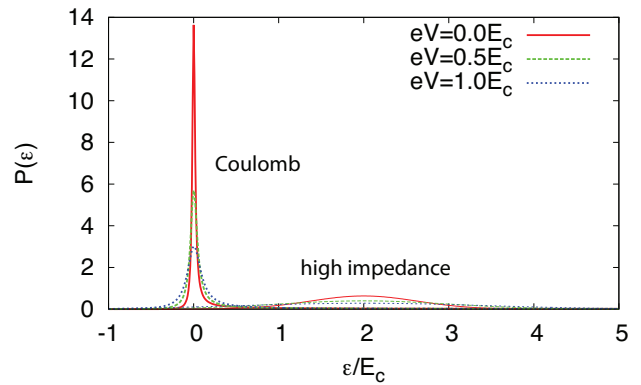}\\
  \caption{(color online) Far-from-equilibrium probability function $P(\varepsilon)$ at different voltages $eV/E_c=0,0.5,1$ for both the high-impedance and the dynamic Coulomb interaction cases. The temperature is $T/E_c=0.1$ and $\alpha=100$, where the parameter $\alpha = e^2 d\nu$ is defined below Eq.~(\ref{tildeU}). Here $E_c$ is the Coulomb energy of the nanojunction.
  }\label{fig.P}
\end{figure}

Next, we discuss the current-voltage characteristics, Eq.~(\ref{eq:I}), of a nano tunnel junction comprised of two thin two-dimensional (2D) disordered conductors (leads) taking into account the effect of Coulomb interaction explicitly. 
To this end we need to calculate the distribution function $P(\varepsilon)$, appearing in the tunneling 
rate, Eq.~(\ref{eq:Gin}) from first principles.

\begin{figure}[t]
\includegraphics[width=0.95\columnwidth]{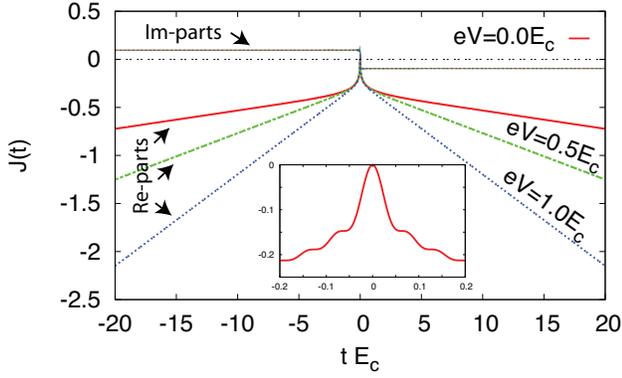}\\
  \caption{(color online) Far-from-equilibrium phase function $J(t)$, Eq.~(\ref{eq.J}), at different voltages $eV/E_c=0,0.5,1$ for the dynamic Coulomb interaction case. The temperature is $T/E_c=0.1$ and $\alpha=100$, where the parameter $\alpha = e^2 d\nu$ is defined below Eq.~(\ref{tildeU}) and $E_c$ is the Coulomb energy of the nanojunction. The real and imaginary parts are plotted separately and the inset shows the behavior of the real part of function $J(t)$ for small times (in the zero voltage case). The latter demonstrates that the high-impedance expansion works in this regime.
  }\label{fig.J}
\end{figure}

In general this function can be written as $P(\varepsilon)=\int_{-\infty}^\infty dt \exp[J(t) + i\varepsilon t]$,
where the function $\exp[J(t)]$ accounts for the interaction with the Bosonic environment.  The far-from-equilibrium function $J(t)$ can be written as~\cite{chtch+prl09}
\begin{equation}\label{eq.J}
\frac{J(t)}{2}=\int\limits_{\tau_e^{-1}}^\infty \frac{d\omega}{\omega}\rho(\omega)
\left[N_\omega e^{i\omega t}+ (1 + N_\omega)e^{-i\omega t}-B_\omega\right],
\end{equation}
where the terms proportional to $N_\omega$ and $1+N_\omega$ correspond to the absorbed and emitted environment excitations, respectively, and
$B_\omega = 1 + 2 N_\omega$. (Here we concentrate on the simplest case when the temperature of absorbed and emitted excitations is the effective electron temperature $T_e$ determined by the environment.) In equilibrium $N_\omega$ reduces to the Bose-function and the functional $P(\omega)$ recovers the result of Ref.~[\onlinecite{Grabert_ch1}]. The energy relaxation time $\tau_e$ in the expression for $J(t)$ determines the low energy cut-off, since the electrons start to equilibrate on larger time scales, i.e. the non-equilibrium description does not hold anymore.

The spectral function $\rho(\omega)$ in Eq.~(\ref{eq.J}) is the probability of the electron--environment interaction. 
We assume that leads are identical and have the same diffusion coefficients $\D^{(L)} = \D^{(R)} \equiv \D$ and densities of states, $\nu^{(L)} = \nu^{(R)} \equiv \nu$. 
For the dynamic Coulomb interaction the spectral function $\rho(\omega)$ can be found, following Ref.~[\onlinecite{rollbuehler+prl01}], as
\begin{equation}
\label{rho2}
\rho_{\mathrm{ij}}(\omega)
=\frac{\omega}{2\pi}\Imm\sum_{\mathbf q}\frac{\left(\frac{2\pi}{L}\right)^2(2\delta_{\rm ij}-1)\tilde U_{\rm ij}({\mathbf q},\omega)}
{(\D q^2-i\omega)^2 }\,,
\end{equation}
where $i,j=1,2$ are the lead indices for the left and right side respectively, and $\tilde U_{\rm ij}({\mathbf q},\omega)$ are the dynamically screened Coulomb interactions within (across) the electrodes.
The form of spectral probability $\rho(\omega)$ [$\rho(\omega)=2\rho_{12}+\rho_{11}+\rho_{22}$] depends on the structure of the environmental excitations spectrum and, thus, on the external bias.

The screened Coulomb interaction in Eq.~(\ref{rho2}) in Fourier space has the form $\underline{\tilde U}(\mathbf{q},\omega) = \{[\underline{U}^{(0)}(\mathbf{q},\omega)]^{-1}+{\underline{\mathcal{P}}}(\mathbf{q},\omega)\}^{-1}$,
where $\underline{U}^{(0)}(\mathbf{q},\omega)=u(q)\underline{I}+v(q)\underline{\sigma}_x$ is the bare Coulomb interaction and $\underline{\mathcal{P}}(\mathbf{q},\omega)$ the
polarization matrix respectively with $\mathcal{P}_{\rm ij}=\nu \D q^2(\D q^2-\imath\omega)^{-1}\delta_{\rm ij}$.

Below we consider quasi-two-dimensional  (2D) infinite leads meaning that $ a < l \ll L$, where $a$ is the lead thickness, $l$ the electron mean free path, and $L$ the 
lead size in the $x$ and $y$ directions. In this case the bare Coulomb interaction has the form
\begin{equation}
U_{ij}^{(0)}(\mathbf{r}_i-\mathbf{r}_j)=e^2\int dz_i\, dz_j\, \frac{\delta(z_i-z^{(0)}_i)\delta(z_j-z^{(0)}_j)}{|\mathbf{r}_i-\mathbf{r}_j|}\,,
\end{equation}
with $z^{(0)}_i=(1/2-\delta_{i1})d$ and $d$ being the nano junction size (distance of the contacts), leading to $u(q)=2\pi e^2/q$ and $v(q)=2\pi e^2 e^{-qd}/q$.

\begin{figure}[t]
\includegraphics[width=0.95\columnwidth]{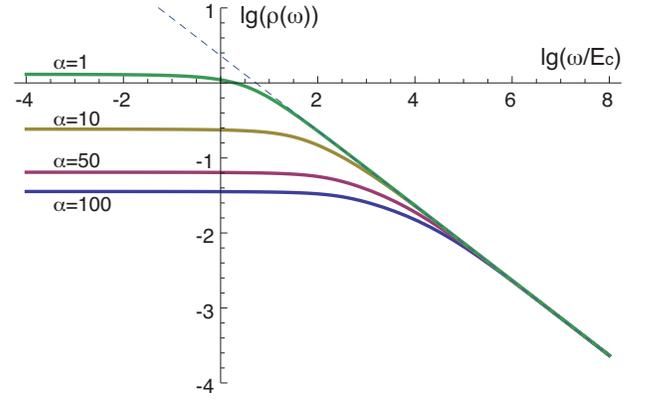}\\
  \caption{(color online) Spectral density function $\rho(\omega)$ for different values of dimensionless parameter 
  $\alpha=1, 10, 50, 100$ in double-log representation. The parameter $\alpha = e^2 d\nu$ is defined below Eq.~(\ref{tildeU}). The data are plotted for the case when Thouless energy, $E_{\rm th}$ is equal to the Coulomb energy, $E_{\rm th} = E_c$.
  }\label{fig.rho}
\end{figure}

The dimensionless matrix elements $\tilde U_{ij}$ of the dynamically screened Coulomb interaction (in units of $e^2 d$) are then given by
\begin{equation}
\tilde U_{\rm ii} = \frac{4 \pi}{\tilde{q}}  \frac{\chi(\tilde{q})}{\chi^2(\tilde{q})- \coth^{-2}(\tilde{q})}\,,\,\,  \tilde U_{\rm i\neq j}=\frac{\tilde U_{ii}}{\chi(\tilde{q})\coth(\tilde{q})}\label{tildeU}
\end{equation}
where $\tilde{q}=dq$ and $\tilde\omega\equiv \omega (d^2/\D)$ with the dimensionless
function $\chi(\tilde{q}) \equiv 1+\coth(\tilde{q})+\frac{4\pi \alpha \tilde q}{ \tilde{q}^2 - i \tilde\omega}$ and $\alpha=e^2 d\nu$.
Using these notations, we can write Eq.~(\ref{rho2}) as
\begin{equation}
\label{rho3}
\rho(\tilde\omega) = \frac{2 e^2 d}{\D} \tilde\omega \Imm  \int\limits_0^\infty \tilde{q} d \tilde{q}  \frac{\tilde U_{11} \left[1-\left(\chi(\tilde{q})\coth(\tilde{q})\right)^{-1}\right]}{(\tilde{q}^2 - i\tilde\omega)^2}\,.
\end{equation}
The spectral function $\rho$ is plotted in Fig.~\ref{fig.rho} as a function of frequency for different values of the
dimensionless parameter $\alpha = 1, 10, 50, 100$. Notably, the $\rho$-function  depends only weakly on frequencies 
in the low frequency limit and decays algebraically as $\rho \sim 1/\omega^{1/2}$ at very high frequencies. Here, we remark that the parameter $\alpha$ has a typical value of $100$ in the metallic case we are considering here.

Using Eq.~(\ref{rho3}) we numerically evaluate the behavior of the $J(t)$-function, Eq.~(\ref{eq.J}), which accounts for the
interaction with the environment. Its behavior for a typical parameter value of $\alpha = 100$ at different electron temperatures is presented in Fig.~\ref{fig.J}. The imaginary part of $J$-function is antisymmetric and almost voltage independent, while the slope of the real part is voltage dependent. 
The imaginary part contributes to the oscillatory factor in the expression for probability function $P(\varepsilon)$ introduced above 
Eq.~(\ref{eq.J}). The behavior of real part is more important, since it describes the interaction with the environment and makes the $P$-integral convergent. 
At small dimensionless times ($tE_c \ll 1$) the real part of $J(t)$ function has quadratic behavior (see the insert in 
Fig.~\ref{fig.J}), which corresponds to the high impedance limit for the environment discussed in detail before.  
However, at larger time scales the $J$-function shows linear behavior, which modifies the probability function $P$ significantly. 

The behavior of $P(\varepsilon)$-function is crucial for calculation of current-voltage characteristics in Eq.~(\ref{eq:I}). We present
the  normalized probability function $P(\varepsilon)$ versus dimensionless energy in Fig.~\ref{fig.P}. In the case of dynamic Coulomb interaction, the $P$-function has two distinct features: i) a peak at low energies and ii) a long tail at high energies. The first feature is related to the fact that in the limit of dynamic Coulomb interaction the screening effects are very pronounced and thus the original bare interaction $E_c$ is completely screened. This is in contrast to the behavior of the $P(\varepsilon)$-function in the high impedance environment, which has a peak at
energies of order of Coulomb energy $E_c$, see Fig.~\ref{fig.P} for comparison. The second feature of probability function $P(\varepsilon)$, the appearance of a long tail, increases the number of available states for energy absorption/emission of the environment enhancing the overall tunneling probability through the junction. This in combination with screening effect results in a significant enhancement of the current at low voltages as compared to the high-impedance case for the metallic value of $\alpha=100$. If parameter $\alpha$ is decreased, i.e. the density of states lower, the Coulomb blockade gets restored, but also the resistance in the Ohmic regime increases. A current-voltage characteristics for $\alpha=50$ is plotted in Fig.~\ref{fig.IVcomp} (dashed, turquois), showing the enhancement of the current at low voltages compared to the high impedance case.

\section{Discussions}

Here we discuss the behavior of probability function $P(\varepsilon)$ in Eq.~(\ref{eq:P1}) and comment on the validity of our approach at 
low temperatures. From Eq.~(\ref{eq:P1}) follows that the probability function $P(\varepsilon)$ for zero temperature is proportional to the  delta-function,  $P(\varepsilon) \sim \delta(\varepsilon - 2E_c)$, meaning that no electron transport is possible below the Coulomb threshold. 
This is a consequences of our consideration of the tunneling transport in Eq.~(\ref{eq:I}) being described in the lowest order in tunneling Hamiltonian. In this approximation higher order effects like electron co-tunneling~\cite{glatz+prb10} is not taken into account. Co-tunneling, introduced in Ref.~[\onlinecite{Averin+prl90}], provides a conduction channel at low applied biases and temperatures, where otherwise the Coulomb blockade arising from electron-electron repulsion would suppress the current flow. The essence of a co-tunneling process is that an electron tunnels via virtual states thus bypassing the huge Coulomb barrier. There are two mechanisms of co-tunneling processes, elastic and inelastic. At very low temperatures only elastic co-tunneling exists meaning that electrons propagate through all virtual states without emitting/absorbing energy.  

In this paper we only consider low (but not very low) temperatures were a bath (phonons) is inefficient (the typical validity temperature range would be between 1K and 100K). Therefore our approach is valid when the interaction time between electrons and many-body excitations (environment) is much smaller than the one between environment and bath, which is the case at not very high temperatures where the number of phonons is small. If the environment interacts strongly with the bath, relaxation is provided by phonons (bath) and $P(\varepsilon) = \delta(\varepsilon)$. In that case Eq.~(\ref{eq:I}) reproduces Ohm's law. 

Last, we comment on two recent experiments of Refs.~[\onlinecite{cohen+prb11,ovadia+prl09}] where 
instabilities in the current-voltage characteristics were observed. In these experiments the properties of macroscopic systems,  arrays of nano-junctions, were studied and therefore our results can not be directly applied to these results. However, transport through a quantum nano-material can be reduced to the single junction problem with an effective medium that plays also the role of a thermostat. This work is currently in progress, but requires in contrast to the single junction also to consider heating effects of the medium~\citep{glatz+prb11}.

In conclusion, we studied the tunneling transport through a nano-junction in the far-from-equilibrium regime at relatively low temperatures. We showed that the current-voltage characteristics is significantly modified as compared to the usual quasi-equilibrium result and demonstrated that for two cases: the high impedance and the dynamic Coulomb interaction case.
One can expect that our results will be important for electron transport in junction arrays, which will be the subject of a forthcoming work.

\begin{acknowledgments}
A.~G. was supported by the U.S. Department of Energy Office of Science under the Contract No. DE-AC02-06CH11357.
I.~B. was supported by an award from Research Corporation for Science Advancement. 
\end{acknowledgments}

\bibliography{noneqIV}

\end{document}